\documentclass[amsmath,amssymb,prl,reprint]{revtex4-1}
\usepackage{graphicx}% Include figure files
\usepackage{dcolumn}% Align table columns on decimal point
\usepackage{bm}% bold math
\usepackage{color}

\setcitestyle{super}

\begin{document}

\title{Noise insights into electronic transport}

\author{S.U.~Piatrusha}
\affiliation{Institute of Solid State Physics, Russian Academy of
Sciences, 142432 Chernogolovka, Russian Federation}
\affiliation{Moscow Institute of Physics and Technology, Dolgoprudny, 141700 Russian Federation}
\author{L.V.~Ginzburg}
\affiliation{Institute of Solid State Physics, Russian Academy of
Sciences, 142432 Chernogolovka, Russian Federation}
\affiliation{Moscow Institute of Physics and Technology, Dolgoprudny, 141700 Russian Federation}
\author{E.S.~Tikhonov}
\affiliation{Institute of Solid State Physics, Russian Academy of
Sciences, 142432 Chernogolovka, Russian Federation}
\affiliation{Moscow Institute of Physics and Technology, Dolgoprudny, 141700 Russian Federation}
\author{G.~Koblm\"{u}ller}
\affiliation{Walter Schottky Institut, Physik Department, and Center for Nanotechnology and Nanomaterials, Technische Universit\"{a}t M\"{u}nchen, Am Coulombwall 4, Garching 85748, Germany}
\author{A.V.~Bubis}
\affiliation{Institute of Solid State Physics, Russian Academy of
Sciences, 142432 Chernogolovka, Russian Federation}
\affiliation{Moscow Institute of Physics and Technology, Dolgoprudny, 141700 Russian Federation}
\affiliation{Skolkovo Institute of Science and Technology, Nobel str. 3, Moscow, Russian Federation}
\author{A.K.~Grebenko}
\affiliation{Moscow Institute of Physics and Technology, Dolgoprudny, 141700 Russian Federation}
\affiliation{Skolkovo Institute of Science and Technology, Nobel str. 3, Moscow, Russian Federation}
\author{A.G.~Nasibulin}
\affiliation{Skolkovo Institute of Science and Technology, Nobel str. 3, Moscow, Russian Federation}
\affiliation{Department of Applied Physics, School of Science, Aalto University, P.O. Box 15100, FI-00076 Aalto, Finland}
\author{D.V.~Shovkun}
\affiliation{Institute of Solid State Physics, Russian Academy of
Sciences, 142432 Chernogolovka, Russian Federation}
\affiliation{Moscow Institute of Physics and Technology, Dolgoprudny, 141700 Russian Federation}
\author{V.S.~Khrapai}
\affiliation{Institute of Solid State Physics, Russian Academy of
Sciences, 142432 Chernogolovka, Russian Federation}
\affiliation{Department of Physics, Moscow State University of Education, 29 Malaya Pirogovskaya St, Moscow, 119435, Russian Federation}

\begin{abstract}

Typical experimental measurement is set up as a study of the system's response to a stationary external excitation. 
This approach considers any random fluctuation of the signal as spurious contribution which is to be eliminated via time-averaging or, equivalently, bandwidth reduction. Beyond that lies a conceptually different paradigm -- the measurement of the system's spontaneous fluctuations. The goal of this overview article is to demonstrate how current noise measurements bring insight into hidden features of electronic transport in various mesoscopic conductors, ranging from 2D topological insulators to individual carbon nanotubes.

\end{abstract}

\maketitle

%$$\mu^L_\uparrow$$
%$$\mu^L_\downarrow$$
%$$\mu^R_\uparrow$$
%$$\mu^R_\downarrow$$
%$$\mu^L_\uparrow$$
%$$\mu_0$$
%$$\mu_\uparrow,\,\mu_\downarrow$$
%$$I_1$$
%$$I_0$$
%$$I_2$$

Back in 1918, in his research for factors limiting the performance of hot cathode amplifiers, Walter Schottky was the first to understand that discreteness of the elementary charge, $e$, gives rise to current fluctuations in a vacuum tube~\cite{Schottky1918}. The noise spectral density is determined by $S_I=2eI$, where $I$ is the tube's average current. Along with the Johnson-Nyquist noise in thermal equilibrium~\cite{Blanter2000}, such fluctuations -- now called the shot noise -- represent one of the two fundamental sources of current noise in a generic conductor.  Careful experiments by Hull and Williams~\cite{Hull1925} confirmed that the shot noise is observable in the current saturation regime of the vacuum tube and provides a way to measure the elementary charge as accurate as in Millikan's oil-drop experiment. Some 70 years later, shot noise experiments demonstrated $e/3$ quasiparticle charge in the fractional quantum Hall effect~(FQHE)~\cite{dePicciotto1997,Saminadayar1997}, one of the milestones in modern physics. 

Diluted electron flow in a vacuum tube obeys Poissonian statistics for purely classical reasons~\cite{Schoenenberger2001}. By contrast, in solid-state conductors, the shot noise arises from the random partitioning of a degenerate electron stream owing to scattering off disorder or inhomogeneities~\cite{Blanter2000}. This results in much richer possible outcomes of the shot noise measurement and brings valuable information about charge transport mechanism, making noise an attractive experimental tool in mesoscopic physics. 

Already in the simplest case of single-mode quantum phase-coherent conductor, the partition noise acquires binomial statistics~\cite{levitov_lesovik1993}. Here, the limits are zero noise in the ballistic case (no scattering) and the full Schottky value in the case of negligible transmission probability, $Tr\ll1$, (tunneling). In between, the spectral density of the current noise is conveniently expressed by a Fano factor $F=S_I/2eI$, which in this case equals~\cite{Reznikov1995} $F=1-Tr$. In multi-mode conductors the current fluctuations in different eigenchannels are independent and the overall $F$ is given by averaging over the eigenvalue distribution~\cite{Blanter2000}. 

In metallic diffusive conductors the transmission eigenvalue distribution is universal~\cite{beenakkerRMP}, i.e. independent of the shape, length and dimensionless conductance, so is the Fano factor $F=1/3$~\cite{Beenakker1992}. In fact,  this universality is even much stronger and persists in the classical limit, given the energy relaxation is negligible on the time scale of diffusion across the device~\cite{Nagaev1992}. That is, perhaps, why in a number of experiments, from normal metals~\cite{Henny1999} to topological insulators~\cite{TikhonovPRL2016} and semiconducting nanowires~\cite{TikhonovSREP2016}, $F\approx1/3$ is reported even though the condition of phase-coherence is apparently not fulfilled. 

Obviously, current noise measurements are capable to highlight the features of electronic transport hidden in conductance experiments~\cite{Blanter2000}. In the ballistic regime, vanishing of the shot noise demonstrates that scattering is suppressed~\cite{Reznikov1995} without performing a bulky length-dependence statistics. In the localized regime, it brings information about the randomness of hopping transport~\cite{Kuznetsov2000,Safonov2003}, which is otherwise encoded in the temperature dependence of the conductance~\cite{Shklovskii_book}. In the diffusive regime, various deviations from $F=1/3$ can arise from inelastic scattering~\cite{Nagaev1992,Nagaev1995,Kozub1995} or due to the presence of built-in tunnel barriers and interfaces~\cite{Nazarov1994}. A broader application list includes measurements of electron-phonon cooling rates~\cite{Roukes1985}, local~\cite{TikhonovSREP2016} thermometry and spectroscopy~\cite{Piatrusha2017}, spin-to-charge conversion~\cite{Meair2011,Arakawa2015,Khrapai2017}, investigation of Coulomb interaction effects~\cite{Roche2004,Yamauchi2011,Harabula2018,Inoue2014,Tikhonov2014} and proximity induced superconducting correlations~\cite{Kozhevnikov2000,Choi2005,Das2012,Ronen2016,TikhonovPRL2016}.
 
This article gives a brief overview of our recent research and is mainly intended to illustrate the strength and, perhaps, the beauty of the noise measurements approach. The body of the paper is divided into five sections, which are largely mutually independent. In section I we investigate the noise in the hopping regime in a quantum {\color{black} Hall}~(QH) insulator. Section II is devoted to the shot noise of the edge transport in HgTe inverted band quantum wells. Section III addresses the problem of interface quality in semiconducting nanowires with metallic contacts. In section IV we study shot noise in a Coulomb blockaded carbon nanotube. Section V gives the estimate of the voltage noise in a resistive state of a superconducting film owing to spontaneous fluctuations of electronic temperature. 

\section{I. Shot noise of a quantum Hall insulator}

\begin{figure}[t]
\begin{center}
\vspace{0mm}
\textsc{\textsc}\includegraphics[width=.9\columnwidth]{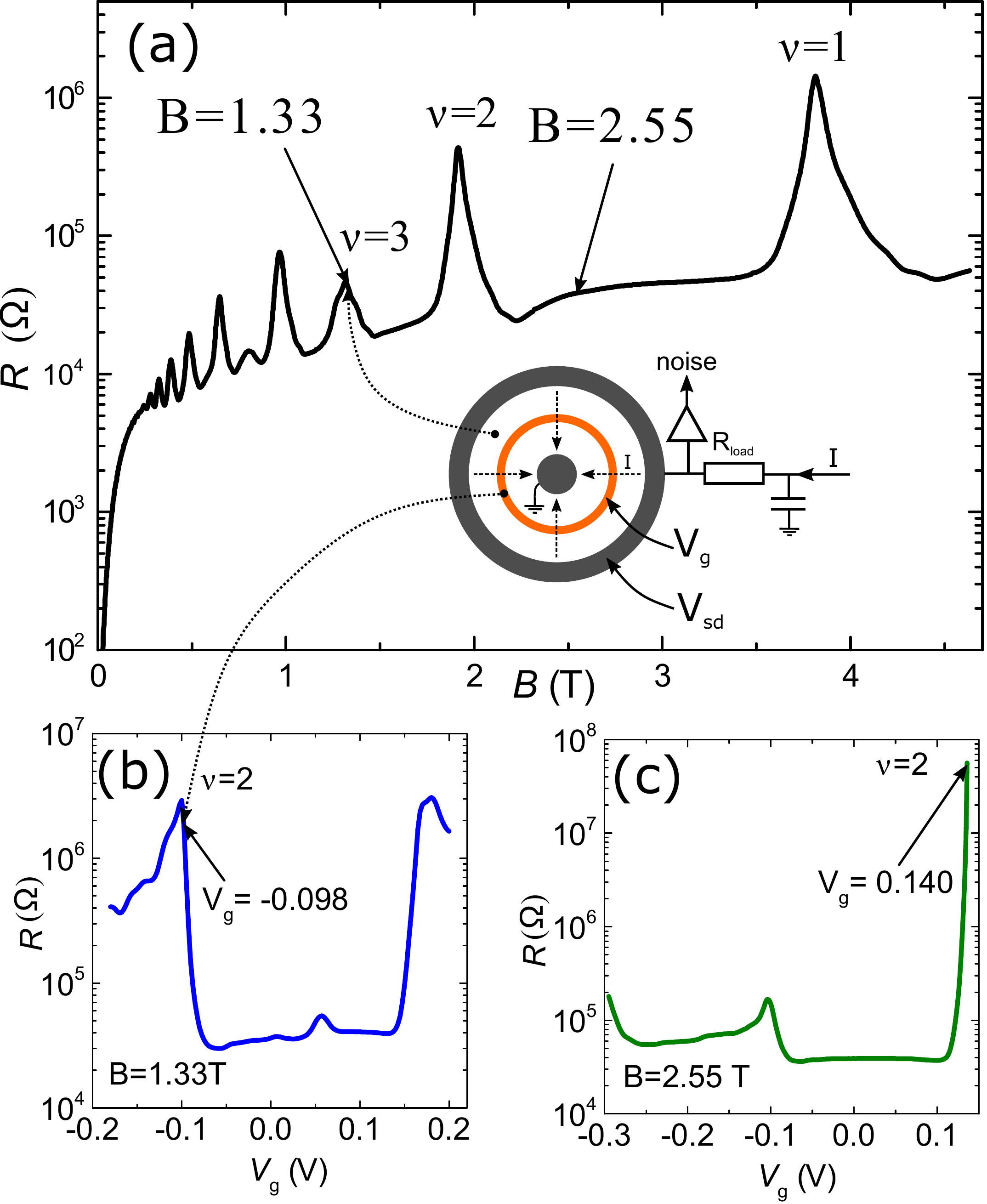}
\end{center}
\caption{Transport measurements and the parameters choice. (a)~Device resistance $R \propto 1/\sigma_{xx}$ as a function of the magnetic field~$B$. Inset: device schematics and the measurement configuration. (b,c)~Dependence of $R$ on the gate voltage~$V_g$ in $B=1.33~\mathrm{T}$~(b) and in $B=2.55~\mathrm{T}$~(c). Arrows mark $V_g$ values used for the noise experiments.}\label{Corbino_fig1}
\end{figure}

In this section, we address the statistics of current flow via insulating states realized in high-quality two-dimensional electron system (2DES) in GaAs in quantizing perpendicular magnetic fields. In the regime of integer quantum {\color{black} Hall} effect~(QHE)~\cite{prange2012quantum}, the Fermi level of 2D electrons falls in the band of localized states and the current preferably flows perpendicular to the electric field. Dissipative transport via localized states is conveniently investigated in the geometry of Corbino disk, see the inset of fig.~\ref{Corbino_fig1}a, which allows one to get rid of the dissipationless Hall current contribution, including edge states.  

Along with exponentially strong  temperature ($T$) dependence of the conductance, a marked transport property of the QH insulating states is its strongly nonlinear current-voltage ($I-V$) response, known as the breakdown of the QHE~\cite{Eber1983}. Although to our best knowledge, there exists no accepted microscopic description of the corresponding $I-V$ characteristics, at least qualitatively one can define two distinct transport regimes. At small enough currents, which we will address below, conduction in the band of localized states occurs via hopping scenario, nearest-neighbor or variable-range, depending on $T$. By contrast, at very high currents, breakdown scenario is realized, during which some electrons gain enough energy to ionize a number of localized electrons, analogous to impact ionization. In this regime, the differential conductance increases by several orders of magnitude accompanied by huge current noise~\cite{Kobayashi2014} with effective Fano factor up to $F\sim10^3$, which is a clear signature of extreme avalanche bunching of electrons.

Here we concentrate on noise at small enough currents when charge transport occurs via individual hopping events within the band of localized states. In few micrometers long devices, the number of hops across the sample can well be a few tens and the noise becomes sensitive to the distribution of their probabilities, or, equivalently, resistances~\cite{Shklovskii_book}. In the regime of variable range hopping~(VRH), this distribution is exponentially wide and both current and noise are dominated by the most resistive hops, so-called hard-hops~\cite{Kuznetsov2000,Safonov2003}. In this case, the Fano factor is inversely proportional to the number of hard-hops, $F\sim1/N$, and for $N=1$ can reach $F\sim1$, as predicted by theory~\cite{Kinkhabwala2006} and consistent with experiments. Remarkably, in GaAs 2DES at low $T$ and deep enough in the insulating regime the Poissonian noise with $F = 1\pm0.1$ was demonstrated~\cite{Tikhonov2013}. This motivates us to study shot noise in quantizing magnetic fields, where Poissonian current statistics would enable, e.g., a direct measurement of the quasiparticle charge in the bulk of the insulating state in the FQHE.

Our device is made of high-quality 2DES of a GaAs/AlGaAs heterostructure with 2D electrons buried 200\,nm below the surface. The as-grown electron density and mobility (at 4.2K) of the 2DES are, respectively, $0.96 \times 10^{11}$~cm$^{-2}$ and $4 \times 10^6 \mathrm{cm^2/Vs}$. On the surface of the structure, we defined a Corbino disk device, schematically shown in fig.~\ref{Corbino_fig1}a. The gate electrode encircles the inner ohmic contact and has the width $L\rm \approx3\,\mu m$ and is 2.15\,mm in perimeter. The distance between inner and outer ohmic contacts is $\rm 120\,\mu m$. All transport measurements were performed using dc excitation in a $^3$He insert and the noise setup is calibrated by Johnson-Nyquist thermometry.

\begin{figure}[t]  
\begin{center}
\vspace{0mm}
\textsc{\textsc}\includegraphics[width=\columnwidth]{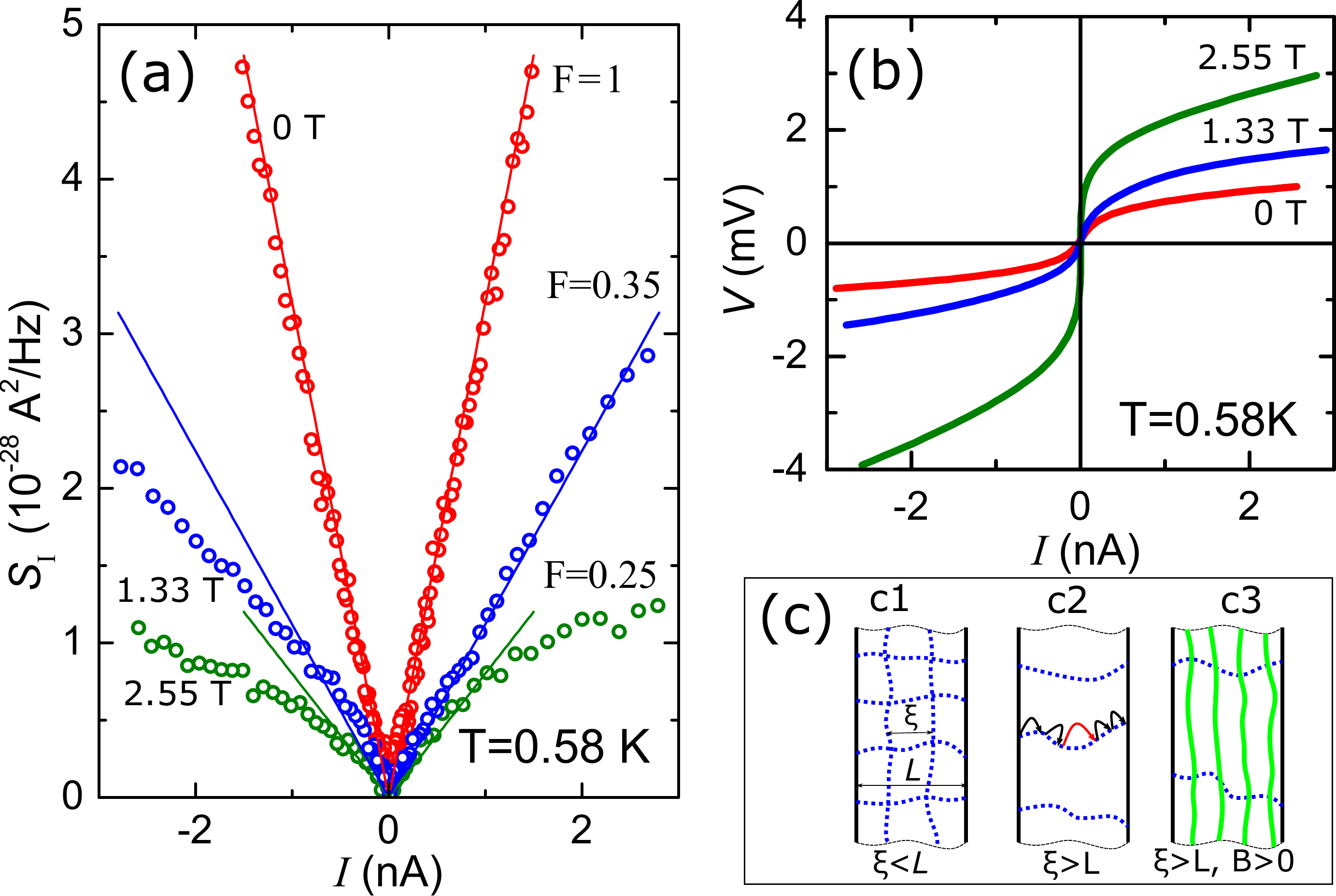}
\end{center}
\caption{Noise and transport in quantized magnetic fields. (a)~Dependence of the noise spectral density~$S_I$ on the current~$I$ in $B = 1.33~\mathrm{T}$ at $V_g =-0.098~\mathrm{V}$ (blue symbols) and in $B = 2.55~\mathrm{T}$ at $V_g =0.140~\mathrm{V}$ (green symbols). For comparison, the data in $B = 0$ at $V_g =-0.313~\mathrm{V}$ are also shown (red symbols). Solid guides correspond to  the fixed Fano factor values of $F=1$,\,0.35 and\, 0.25. (b)~$I-V$ characteristics of the device measured simultaneously with the data in panel~(a). Note that the current densities used in our experiment are about two orders of magnitude smaller compared to Ref.~\cite{Kobayashi2014}. (c) Sketches of the current paths in a random resistance network in a long~(c1) and short~(c2) sample in zero magnetic field and in a short sample in quantizing field~(c3).}\label{Corbino_fig2}
\end{figure}
\par

Figure~\ref{Corbino_fig1} illustrates the procedure of the parameter choice for the noise measurements. The $B$-field is chosen so that the Landau level filling factor in the bulk is somewhat below or above $\nu=2$. In these cases, see the arrows in fig.~\ref{Corbino_fig1}a, the Fermi level falls in the band of delocalized states (note still hardly resolved $\nu=3$ spin gap at $B=1.33$\,T) and the resistance of the device at $V_g=0$ does not exceed about 50\,k$\rm\Omega$. After that, we vary the gate voltage thus tuning the filling factor under the gate close to the integer value $\nu=2$, figs.~\ref{Corbino_fig1}b and~\ref{Corbino_fig1}c. Here, the cyclotron gap opens in the 2DES under the gate and the resistance increases by at least two orders of magnitude, with exponentially strong $T$~dependence. 

In figs.~\ref{Corbino_fig2}a and ~\ref{Corbino_fig2}b we plot, respectively, the dependencies of shot noise and voltage across the device on the current for the chosen combinations of $B$ and $V_g$. For comparison, the data obtained in the absence of magnetic field in the regime of hopping conduction at $V_g =-0.313~\mathrm{V}$ are also shown. All the $I-V$ curves in fig.~\ref{Corbino_fig2}b are strongly nonlinear, which is typical for conduction via localized states~\cite{Andrei1988,Jiang1990,Shashkin1994}, and become more resistive at increasing $B$. For instance, the linear response resistance is almost two orders of magnitude higher in $B=2.55$\,T than in $B=0$. However, the trend in the shot noise is opposite. As seen from fig.~\ref{Corbino_fig2}a, at increasing magnetic field the shot noise gradually drops from the Poissonian value $F=1$ in $B=0$ to $F\sim0.25$ in $B=2.55$\,T. The finding of shot noise reduction in the insulating states in quantizing magnetic fields is the central result of this section. We briefly discuss it below. 

The observation of $F=1$ in $B=0$ reproduces our previous result~\cite{Tikhonov2013} and signals that transport in zero field
occurs via VRH conduction in the regime of finite-size effect, when the width of the gate, $L$, is not greater than the size of the critical cluster, $\xi$. Compared to longer devices with $L\gg\xi$ in which the random resistance network is very well interconnected~\cite{Shklovskii_book}, see the sketch {\color{black} c1} of fig.~\ref{Corbino_fig2}c, in the finite-size regime the network splits into a number of quasi-1D hopping chains. {\color{black} They provide independent current paths connected in parallel~\cite{Rodin2011}, see the sketch c2 of fig.~\ref{Corbino_fig2}c}. In a magnetic field, naively, one would expect that the finite-size effect further strengthens, for $\xi\propto g^{-7/9}a^{-5/9}$ and both $g$, the density of states at the Fermi level, and $a$, the localization radius, are expected to decrease at increasing $B$. In principle, there are two possibilities to explain the observed reduction of the shot noise by a factor of~$\sim3$. The first, and fantastic, possibility would be the reduction of the quasiparticle charge in the 2DES owing to the fractional correlations~\cite{dePicciotto1997,Saminadayar1997}. This scenario, however, is highly unlikely since FQHE in similar structures is very weak in such small magnetic fields~\cite{Prokudina2008} and no signatures of the fractional gaps are seen in Fig.~\ref{Corbino_fig1}a. The second possibility is that the process of thermal activation of electrons towards delocalized states starts to contribute to transport in quantizing magnetic field. This might result from the $B$-driven suppression of the conductance of VRH network, $\ln{G}\propto -g^{-1/3}a^{-2/3}T^{-1/3}$. Note, however, that in the QH regime thermally delocalized electrons cannot directly take part in dissipative transport since they drift long distances perpendicular to the electric field~\cite{Iordansky1982}. Nevertheless, a variety of choices for the next hop, which emerges in the course of this drift, favors the escape of electrons from the most resistive pathways. In this way, quasi-1D hopping chains interconnect under the gate, see the sketch {\color{black} c3} of fig.~\ref{Corbino_fig2}c, and hopping becomes less random thereby reducing the shot noise. Our measurements demonstrate that reaching the Poissonian current statistics in quantizing magnetic fields is much more challenging compared to the $B=0$~case. Technically, this makes the possibility of the direct quasiparticle charge measurement in the bulk of a QH insulator extremely difficult. 

\section{II. The noise of edge channels in a 2D topological insulator}

The recently introduced topological insulators are a class of materials with an insulating bulk and conducting helical surface states \cite{RevModPhys.82.3045}. The core property of these states is the direct correspondence between electron spin and momentum, called spin-momentum locking, which suppresses elastic disorder scattering by the angle of $\pi$. For 2D topological insulators, where the bulk material is a quantum well (QW) \cite{Bernevig2006,Konig2007,Knez2011} or a monolayer crystal \cite{Qian2014,Wu2018} the surface states represent a 1D single-mode
edge conduction channel, known as helical edge states. In this case, the spin-momentum locking strictly prohibits backscattering 
unless some spin-flip mechanism is present, the property known as topological protection. As a consequence, in the absence of time-reversal symmetry breaking, only inelastic (phase-incoherent) backscattering is allowed, which (i) dictates quantized conductance $G=G_q\equiv e^2/h$ in the limit of $T\rightarrow0$ and (ii) precludes carrier localization, which is inevitable in conventional quasi-1D conductors with a disorder. On the experimental side, the edge transport is clearly demonstrated~\cite{Roth2009,Nowack2013,Olshanetsky2015} and, in the best devices, a poorly quantized ballistic conductance of $G\approx G_q$ is observed over a few micrometer length scale~\cite{Konig2007,Olshanetsky2015,Dantscher_2017,Li_Du_2017}. For longer edges, roughly linear length dependence of the resistance is reported \cite{Gusev2014,Wu2018} with $G_g/G\propto L$, reminiscent of phase-incoherent diffusive transport. Overall, experiment leaves a certain degree of doubt on the strength of the topological protection, especially in view {\color{black} of the} edge transport findings in the trivial phase of InAs/GaSb QWs \cite{Nichele2015}.

\begin{figure}[t]
	\begin{center}
		\includegraphics[width=\linewidth]{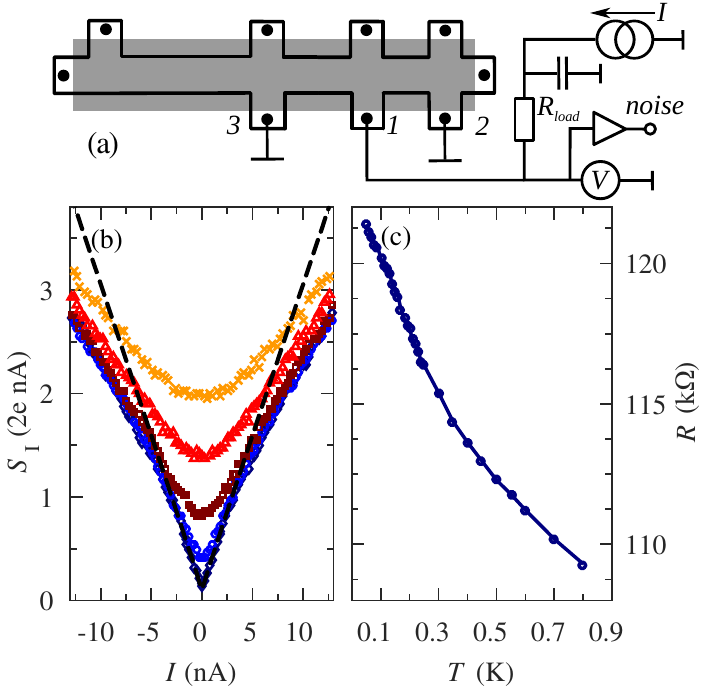}
	\end{center}
	\caption{(a) Device schematic and measurement configuration, including load resistance, DC and noise measurement circuits connected to contact 1. (b) Shot noise measurements versus bias current. The symbols represent shot noise, measured in configuration (a) at temperatures (top to bottom) 800\,mK (crosses), 600\,mK (triangles), 400\,mK (squares), 200\,mK (hexagons) and 80\,mK (diamonds). The dashed line is the shot noise prediction with $F=0.29$ at 80\,mK. (c) Two-terminal linear-response resistance of a 5\,$\rm\mu m$ long edge 1-2 as a function of temperature obtained in a different cooldown.} 
	\label{hgte_fig1}
\end{figure}
 
%Several models have been proposed, trying to combine the presence of helical edge states and $G \ll G_q$, which requires a presence of some spin flip mechanism. This models involve spin relaxation in the charged puddles located in the bulk \cite{Vayrynen2013,PhysRevB.94.045425}, spin-flip via interaction with nuclear momentum \cite{Hsu2017} and spin-flip on the magnetic impurities in the proximity of the edge states \cite{Kurilovich2016,Kurilovich2017}. However, the weak temperature dependence of the resistance, which is commonly observed experimentally \cite{Konig2007,Gusev2014,Tikhonov2015}, is challenging for most proposed models. Concluding, there are two existing problems, concerning 2D TI: the separation between the edge conduction of topologically nontrivial and trivial origins, and (in the nontrivial case) the determination of the dominant backscattering mechanism. None of this problems can be solved, while relying solely on resistance measurements, unless a reproducible length independent $G = G_q$ system would be obtained.

Non-equilibrium noise measurements in the regime of edge transport provide additional information about electron scattering and help, at least, to narrow down the list of possibilities. {\color{black} In Ref.~\cite{Tikhonov2015}  some of us studied 8\,nm wide HgTe QWs in the disordered limit $G \ll G_q$ and measured $0.1 < F < 0.3$. This didn't match a well-known~\cite{Blanter2000} phase-coherent 1D single-mode result $F=1-Tr\approx 1$, where $Tr\equiv G/G_g$ is the transmission probability.} However, in that work, the conclusion about the trivial origin of the edge transport was based on the assumption of dephasing mechanisms not related to spin, which is not the only possibility~\cite{Aseev_2016}. The opposite limit was addressed in Ref.~\cite{Piatrusha2017}, which considered the shortest possible edges with $G \approx G_q$ realized in lateral p-n junctions of a 14\,nm wide HgTe QWs. The observed shot noise is analyzed considering the contact leads overheating. While neither helical edge states nor diffusive multi-mode scenarios were excluded due to unknown hole-phonon coupling in p-type conduction region, the former scenario was found more consistent.

Here we report measurements of the shot noise in the 14nm HgTe QWs. Transport measurements in similar samples were performed in Ref.~\cite{Olshanetsky2015} and noise measurements in p-n junctions in Ref.~\cite{Piatrusha2017}. Samples were grown by molecular beam epitaxy, before being shaped in form of multiterminal hall-bars by wet etching and covered with a 200\,nm thick $\rm SiO_2/Si_3N_4$ insulating layer~\cite{Kvon2011}. On the top of the insulating layer, an Au/Ti metallic gate was deposited, covering the entire hall-bar region. In our devices the edge lengths varied between 2 and 30 $\mu m$, with corresponding conductances in the range $0.3 G_q\gtrsim G\gtrsim 0.05 G_q$. The measurements were performed in a Bluefors dilution refrigerator at electronic temperatures {\color{black} from $80$\,mK to $0.8$\,K},
in order to exclude thermally activated bulk conduction (bulk band gap of $\approx 3$\,meV reported in Ref.~\cite{Olshanetsky2015}). 
In the present experiment, the contact 1 in fig.~\ref{hgte_fig1}a is used to drive current through the two resistive edges 1-2 and
1-3 with lengths 5\,$\rm \mu m$ and 10\,$\rm \mu m$ respectively, connected in parallel. We measure both the shot noise and the DC
voltage from contact 1. 

Figs.~\ref{hgte_fig1}b and ~\ref{hgte_fig1}c are the main results for this section. In the panel (b) the shot noise spectral density is plotted as a function of the device bias current $I$ at different $T$. At the lowest $T=80$\,mK and $|I|<5$\,nA we observe a clear linear shot noise behavior with a Fano factor $F\approx0.29$, see the guide line, very close to the universal diffusive value of 1/3. At increasing $|I|$ and/or $T$ the current noise gradually deviates down from the guide line and the effective Fano factor decreases, e.g. $F \approx 0.09$ at $T=0.8$\,K. At the same time, see fig.~\ref{hgte_fig1}c, the linear response resistance, $R$, exhibits a weakly insulating temperature dependence increasing roughly by 10\,\% at decreasing $T$ from 0.8\,K to 80\,mK. In the following, we briefly discuss these results in the context of possible transport scenarios.

On the one hand, the observation of shot noise Fano factor very close to the universal value $F=1/3$ in diffusive conductors~\cite{Nagaev1992,Beenakker1992} evidences diffusive edge transport in our devices, with negligible energy relaxation. On the other hand, that diffusive transport is observed for edge conductances much smaller than $G_q$, obviously, implies strong dephasing, regardless of the edge transport scenario~\cite{beenakkerRMP,DEJONG1996219}. In the case of trivial multi-mode edge conduction, slightly insulating $T$-dependence of the resistance in fig.~\ref{hgte_fig1}c can  be attributed to a weak localization correction in the limit of dephasing length smaller than localization length. In this case, the correction to the universal value $F=1/3$ is still expected to be small, yet positive~\cite{Blanter2000}. In the case of helical edge states, our data allow to exclude {\it e-e} scattering  as a relevant backscattering mechanism in generic helical liquids~\cite{Schmidt2012,Kainaris2014}. In this situation, one would expect metallic $T$ dependence of the resistance and Fano factor~\cite{Nagaev1995,Kozub1995} $F>1/3$, which is opposite to our experiment. The microscopic model of random charge-puddles coupled to helical edge states~\cite{Vayrynen2013} is inconsistent with the $R(T)$ data in fig.~\ref{hgte_fig1}c. However, a more general phenomenological model of Ref.~\cite{Aseev_2016} captures our noise data correctly, provided negligible energy relaxation and sufficient spin-flip rate within a puddle. In order to explain the behavior of $R(T)$, however, an increase of the spin-flip rate at decreasing $T$ is necessary, which might be a problem for theory in our temperature range. Scenarios of time-reversal symmetry breaking owing to  magnetic impurities~\cite{Kurilovich2017} or hyperfine interaction~\cite{Hsu2017,Hsu2018} predict negative corrections to the conductance without energy relaxation to the bath, which would be qualitatively consistent with our data. Yet, the numerical estimates of these effects in HgTe QWs are unrealistically small~\cite{Kurilovich2017,Hsu2018}.

\section{III. Noise probes interface transparency in hybrid structures}

The theoretical possibility to realize hybrid topological materials~\cite{0034-4885-75-7-076501} has recently led to the revival of experimental interest in inducing superconducting correlations into semiconducting materials. Currently, the most relevant direction is proximitizing quasi-one-dimensional nanowires (NWs) either grown individually~\cite{Mourik1003,Das2012,PhysRevB.87.241401} or realized in 2D electron gas~\cite{PhysRevB.93.155402,PhysRevApplied.7.034029,PhysRevLett.119.176805}. For all possible applications, the quality of both the one-dimensional conductor and of the semiconductor-superconductor interfaces are of paramount importance. In this section, we demonstrate how noise measurements may complement simple transport measurements in order to better track the possible influence of any present potential barriers.
\par
To illustrate this idea we present experimental data obtained on individually grown semiconductor NWs with Al interfaces. Unintentionally doped catalyst-free InAs NWs were grown by MBE on a Si(111) substrate~\cite{0957-4484-21-36-365602}. The NWs were then detached from the substrate in an ultrasonic bath and drop-casted on a piece of commercial $n^+$-doped Si wafer covered by $300\,\mathrm{nm}$ of SiO$_2$. In order to remove the native oxide layer from the contact regions, prior to the electron beam evaporation of Al otherwise identical samples NW1 and NW2 were in-situ exposed to Ar ion gun ($4.5$ sccm Ar flow, $40\,\mathrm{V}$ discharge and
$400\,\mathrm{V}$ accelerating voltage) during $2$ and $4$~minutes, respectively. The latter time is presumably too large and might have led to over-etching of the device NW2 under electrodes and, thus, to bad Al/InAs interfaces. The evaporation was performed in a Plassys MEB 550S system. The measurements were performed in a dry Bluefors dilution refrigerator with the electronic temperature of $\approx120\,\mathrm{mK}$, verified by noise thermometry. 
%The back gate of the device NW2 was leaking at gate voltages larger than $+12\,\mathrm{V}$ and did not permit a wider range tuning of the conduction.
\par
\begin{figure}[h]
	\begin{center}
		\includegraphics[width=0.9\linewidth]{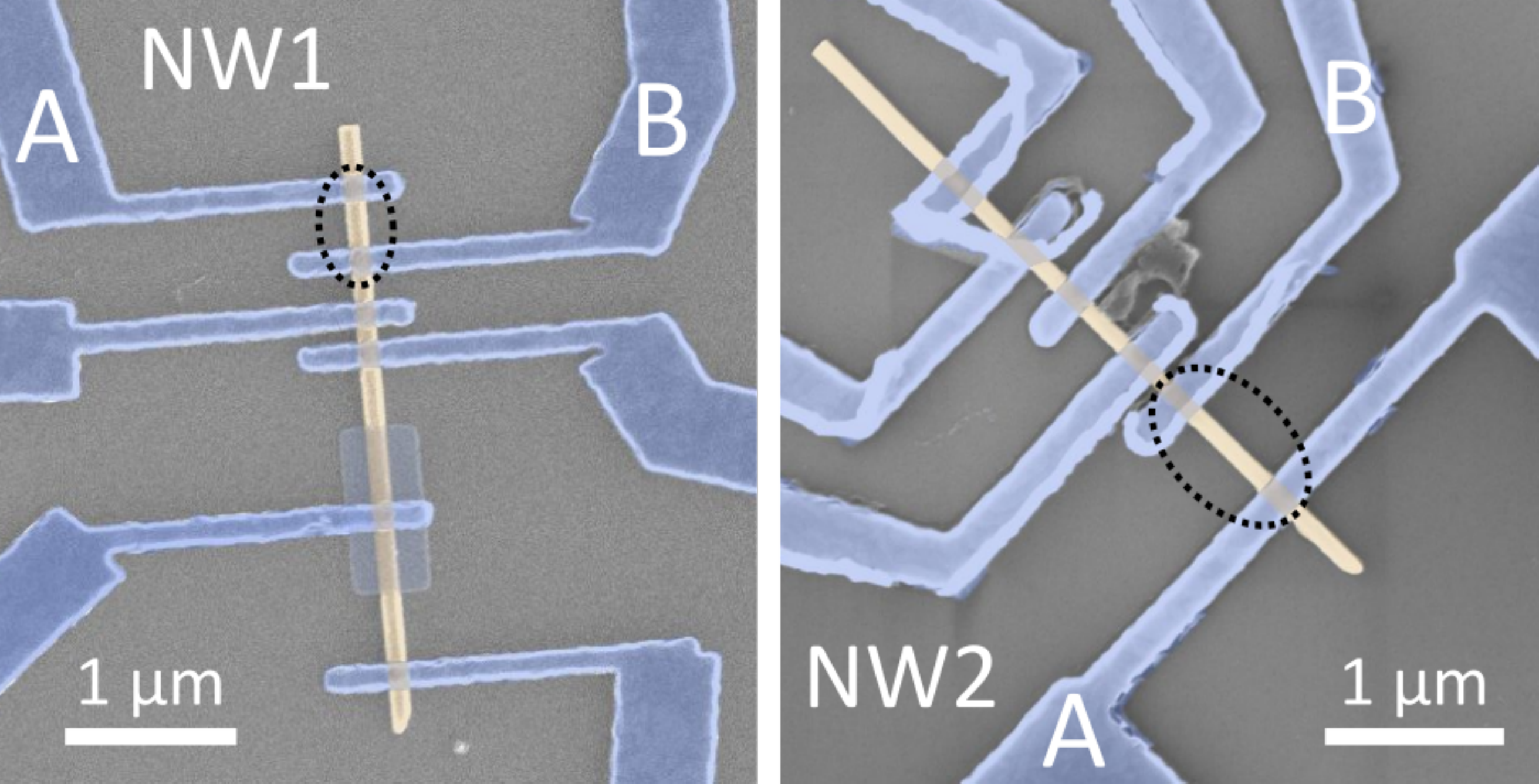}
	\end{center}
	\caption{False color SEM images of the studied devices. The NW sections between contacts A and B (dashed ovals) were used for transport and noise measurements.} 
	\label{NW_fig1}
\end{figure}
\par
False color SEM images of the studied devices are shown in fig.~\ref{NW_fig1}. The Al contact electrodes are nominally $100\,\mathrm{nm}$-wide. We will discuss the data obtained on the NW sections between electrodes A and B, which are $300\,\mathrm{nm}$ and $700\,\mathrm{nm}$ long for devices NW1 and NW2, respectively.
\par
\begin{figure}[h]
	\begin{center}
		\includegraphics[width=0.9\linewidth]{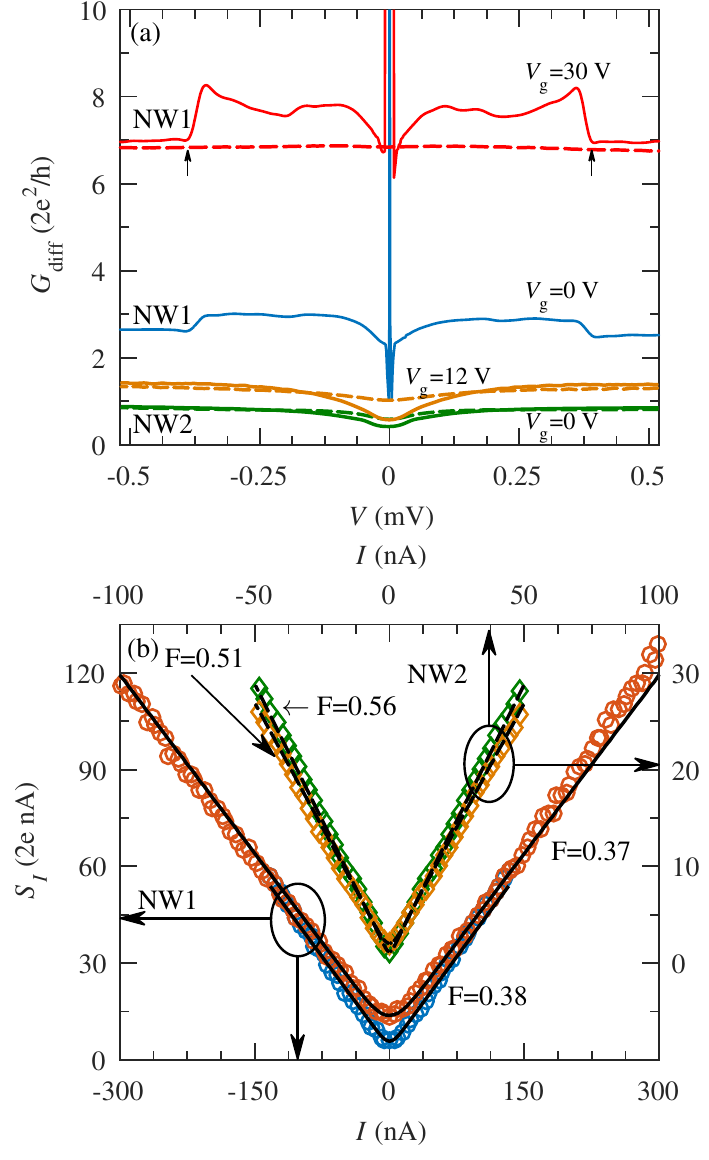}
	\end{center}
	\caption{(a)~Differential conductance against bias voltage for two devices NW1 and NW2 at various gate voltages. Solid and dashed lines are obtained in $B=0$ and $B=120\,\mathrm{mT}$, respectively. (b) The corresponding noise spectral current density in $B=120\,\mathrm{mT}$ at electronic temperature~$T=120\,\mathrm{mK}$. {\color{black} The data for NW1 and NW2 are built, respectively, in left-bottom and right-top axes.}.} 
	\label{NW_fig2}
\end{figure} 
In fig.~\ref{NW_fig2}(a) we plot the bias dependence of the differential conductance, $G=\mathrm{d}I/\mathrm{d}V$, both in zero, $B=0\,\mathrm{T}$ (solid lines), and non-zero, $B=120\,\mathrm{mT}$ (dashed lines), magnetic fields for various back gate voltages~$V_{\mathrm{g}}$. In the shorter device NW1 we observe the supercurrent at $V_{\mathrm{g}}=30\,\mathrm{V}$ and the large zero-bias conductance peak at $V_{\mathrm{g}}=0\,\mathrm{V}$. In fig.~\ref{NW_fig2}(a) these features are obtained by numerical differentiation of the $I-V$-curves, while the larger-bias data are obtained by the lock-in measurements with ac current modulation of $1.5\,\mathrm{nA}$ rms. 
\par
The comparison of traces obtained in superconducting and normal states may indicate the better quality of Al/InAs interfaces in the NW1 device compared to the NW2 device. Indeed, in NW1 in $B=0\,\mathrm{T}$ we observe the increase of $G$ at sub-gap voltages, starting abruptly at approximately twice the superconducting energy gap of Al: $|V|\approx2\Delta/e\approx380\,\mu \mathrm{V}$, see vertical arrows. Microscopically, this sub-gap enhancement above the normal-state conductance~$G_N$ reflects the Andreev reflection charge transport taking place at Al/InAs interfaces and is distinctive for transmissive SNS structures~\cite{PhysRevB.89.214508,0268-1242-32-9-094007}. The observation of the supercurrent/large zero-bias conductance~$G_S=G(\mathrm{V=0,B=0})$ further suggests the good quality of the interfaces in the device NW1. We note that these features remain in NW1 at lower gate voltages when the normal-state resistance of the device is on the order of~$2e^2/h$. On the contrary, in NW2, in the sub-gap region at bias voltages $|V|\lesssim100\,\mathrm{\mu V}$, we observe the decrease of the conductance below the normal-state value in all the available range of~$V_{\mathrm{g}}$. Such a behavior is inherent in nanostructures of a tunnel type~\cite{PhysRevB.87.241401} and reflects the fact that Andreev reflection requires two transmission events~\cite{Shelankov}. The normal-state conductance is thus higher being proportional to the first power of transmission eigenvalues. 
%We note that the change from the normal-state values~$G_N$ is not greater than $\approx0.25G_{N}$ in NW1 and $\approx0.55G_N$ in NW2. These values are intermediate. 
%While in the ballistic case one would expect the conductance doubling in the case of perfectly transmissive interfaces (for NS, ? for SNS) and 
%We note that the maximal changes in differential conductance are 0.4 and 0.55 in respect to the normal-state value.
\par
%forjetpreviewnoise.m
Altogether, the data of fig.~\ref{NW_fig2}(a) display the difference between the devices NW1 and NW2. This difference is additionally manifested in its shot noise in the normal state {\color{black} in magnetic field}. In fig.~\ref{NW_fig2}(b) we demonstrate the shot noise of the devices NW1 (left-bottom axes) and NW2 (right-top axes, the scales on both axes are enlarged by a factor of~$3$ and the data are vertically offset by~$2e\cdot10\mathrm{nA}$ for clarity) obtained in $B=120\,\mathrm{mT}$ at the same values of $V_{\mathrm{g}}$ as in fig.~\ref{NW_fig2}(a). For both samples we observe linear $S_I(I)$-dependencies characteristic of elastic transport. In NW1, the Fano-factor $F\approx0.37$ is almost independent of $V_{\mathrm{g}}$ and is only slightly greater than the universal value $F_0=1/3$, characteristic for diffusive elastic {\color{black} conduction mechanism}. The shot noise in NW2 is significantly greater with $F\gtrsim0.5$, slightly increasing with increasing resistance in the available gate voltage range. In the following we discuss the device NW2 since for NW1 we cannot reliably exclude the influence of experimental uncertainty. In principle, the conductance diminishing in the sub-gap region and the accompanying enhanced shot noise in the device~NW2 might result from the possibly present tunnel barriers~\cite{DEJONG1996219}. These might be caused by the short-period wurtzite segments in our polytypic InAs NWs~\cite{PhysRevB.97.115306} or, alternatively, represent the barriers in the vicinity of Al/InAs interfaces, which result from disorder or band bending due to change in electrostatics after {\color{black} deposition of}  Al~\cite{Deng1557}. 
\par
The Fano-factor of the phase-incoherent diffusive conductor with planar tunnel barriers~$\Gamma_i$ is given by
\begin{equation*}
F=F_0\left(1+2\frac{\sum\limits_{i=1}^{n} R_{\Gamma_i}^3}{\left(\sum\limits_{i=1}^{n} R_{\Gamma_i}+R_D\right)^3}\right),
\end{equation*}
where $R_{\Gamma_i}$ is the resistance of the $i$th barrier, $n$ is the number of barriers and $R_D$ is the sum of the resistances of diffusive pieces. According to this expression, the experimental observation of $F\gtrsim0.5$ in NW2 then necessarily requires that the resistance of the device is dominated by the resistance of a single tunnel barrier~$\Gamma$ with $R_{\Gamma}\gtrsim2R_D$. Were it the interfacial barrier, we can roughly estimate the ratio of the normal-state and zero-field zero-bias conductances using the result of Nazarov~\cite{PhysRevLett.73.1420}. Numerically, we find $G_N/G_S\approx1.7$ and $2.1$ at $V_{\mathrm{g}}=12$ and $0\,\mathrm{V}$, respectively, which is in reasonable agreement with the experimentally observed ratios of $1.7$ and~$1.4$. While this may point in favor of the interfacial barrier, we emphasize that our data also don't exclude the structural defect scenario. {\color{black} Concluding this section, we note that the data presented here is, to our best knowledge, the first attempt  to verify the role of the built-in tunnel barriers from simultaneous transport and noise measurements in diffusive semiconductor-superconductor hybrid structures.}

\section{IV. Shot noise in carbon nanotube quantum dots}

The randomness of electronic transport in mesoscopic conductors in many cases can be understood within the framework of non-interacting electrons~\cite{Blanter2000}. At first glance, this may seem surprising since {\color{black} in the absence of interactions electrons must be uncorrelated and obey Poissonian statistics}, just like in Schottky's vacuum tube. In reality, degenerate electrons in solids are extremely correlated owing to the Pauli exclusion principle~\cite{levitov_lesovik1993}, which is the ultimate reason for the shot noise vanishing in a ballistic conductor~\cite{Reznikov1995}. For the same reason, intrinsic ordering of the incident flux can be directly observed in cross-correlation experiments with electronic beams in the QH edge channels~\cite{HennyScience1999}. On top of this, additional correlations can arise thanks to the proximity with a superconductor~\cite{Blanter2000} or due to Coulomb interactions between electrons. In the latter case, which we focus on in this section, such effects are most pronounced in quantum dots (QDs), where Coulomb energy represents the largest energy scale.

%The simplest (two-terminal) quantum dot is a conductive island weakly coupled to macroscopic leads via tunnel barriers. Here, we investigate  a carbon nanotube (CNT) QD with the leads made of { \bf  MATERIAL?}. The CNT was grown by {\bf method, shortly processing details}. Unlike in lateral quantum dots formed, e.g. in GaAs, tunnel barriers here naturally result from the Schottky barrier between the CNT and the contact metal and the chemical potential of the dot is tuned by a back gate votage $V_{\mathrm{bg}}$. The scanning electron micrograph of the device shown in Fig.~\ref{CNT_fig1}a. On top of the CNT (faint line) three metallic contacts, 300\,nm wide are evaporated, dividing it into two sections, each approximately 400\,nm long. Throughout the experiment, the left most contact was grounded, the rightmost floating and the bias voltage and center contact was biased ?? I-V converter, noise calibtated by J-N. etc. We have checked that Coulomb coupling between the two dots on  either side of the central contact is negligible, hence possible charging events on the right QD are irrelevant for the transport and noise measured in the left QD. All the data below are obtained at a bath temperature of 0.5\,K in a liquid 3$^$He insert.

\begin{figure}
	\centering
	\includegraphics[width=\columnwidth]{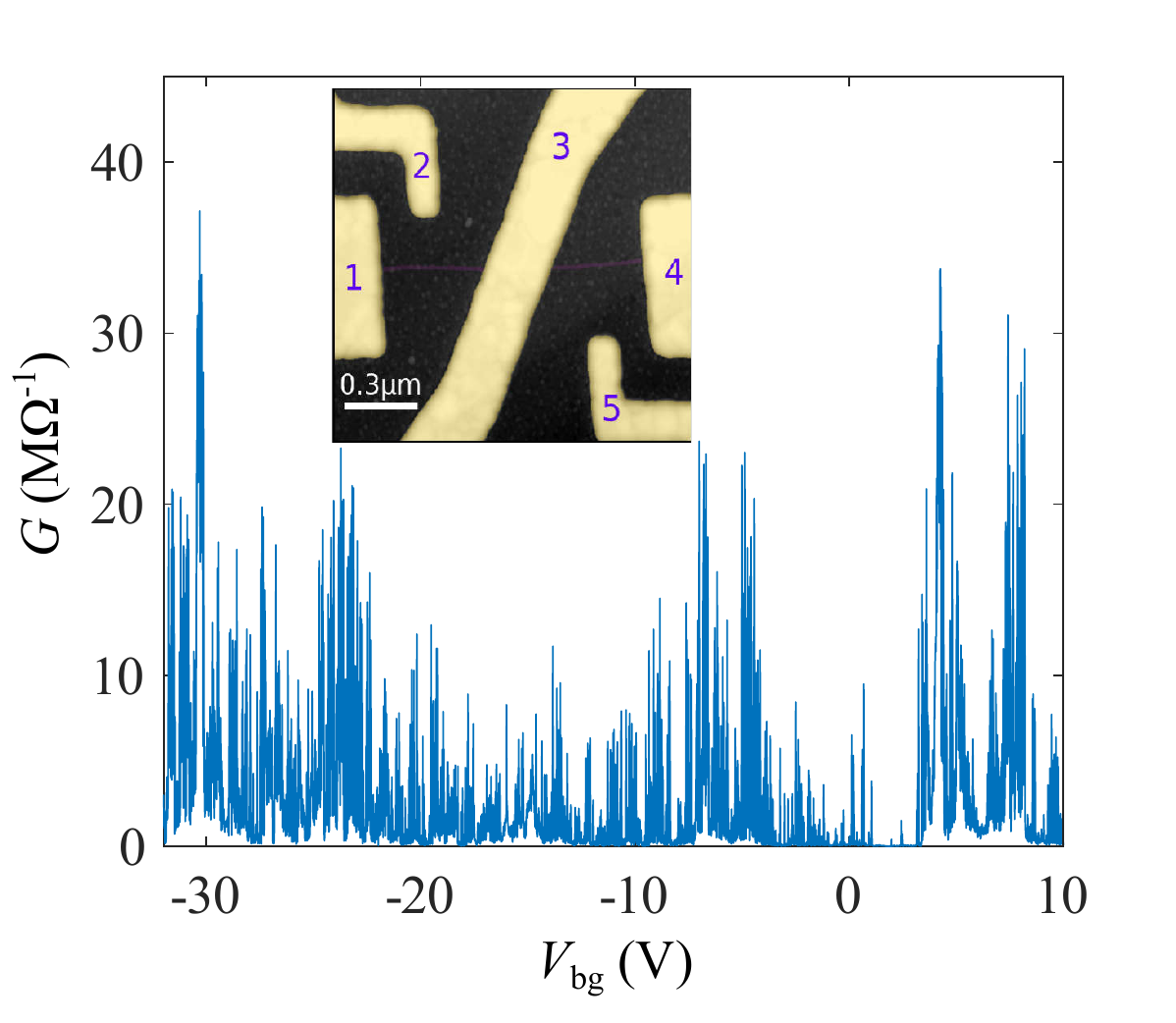}
	\caption{Differential conductance as a function of back-gate voltage at zero applied bias. Inset: atomic force microscopy image (false colors) of the sample. Data discussed in this section were obtained for the left segment of the SWCNT (between metal contacts 1 and 3).}\label{CNT_fig1}
\end{figure}

The simplest (two-terminal) quantum dot is a conductive island weakly coupled to macroscopic leads via tunnel barriers. Here, we investigate a single-walled carbon nanotube (SWCNT) QD with thermally evaporated Ti/Pd/Au leads with thicknesses of, correspondingly, 0.3/5/60\,nm and fabricated via standard e-beam lithography technique. The SWCNTs were grown by aerosol CVD process~\cite{Moisala2006,Tian2011} and dry deposited on a Si/SiO\textsubscript{2} substrate. Unlike in lateral quantum dots formed, e.g., in GaAs, here, tunnel barriers naturally result from the Schottky barrier between the {\color{black} SWCNT} and the contact metal. The chemical potential of the dot is tuned by the back gate voltage~$V_{\mathrm{bg}}$. The atomic force micrograph of the device is shown in fig.~\ref{CNT_fig1} (inset). Nanotube's height measured before all fabrication steps was approximately 1.3\,nm, which to best of our knowledge is an indication of a bundle of SWCNTs, since the minimum diameter of a single nanotube we observed was in the range of 0.6 - 0.8\,nm. On top of the {\color{black} SWCNT} (faint line) three metallic contacts are evaporated, dividing it into two sections, each approximately 400\,nm long. Throughout the experiment, the center contact 3 was grounded, the rightmost contact 4 was floating and the dc bias voltage with a small ac modulation was applied to the left contact 1 via the input of a home-made {\it I-V} converter ($10^7$\,V/A). The current was measured using standard lock-in technique at 31\,Hz modulation frequency. The resonant tank-circuit with the central frequency of $\rm 20.9\,MHz$ and a $5\,\rm k\Omega$ load resistor was connected to the {\color{black} contact 1} for the shot noise measurements. As usual, the signal was calibrated using Johnson-Nyquist noise thermometry.  We used side gates 2 and 5 to check that Coulomb coupling between the two dots on either side of the central contact 3 is negligible, hence possible charging events on the right QD are irrelevant for the transport and noise measured in the left QD. All the data discussed below are obtained at a bath temperature of 0.5\,K in a liquid $^3$He insert with grounded gates 2 and 5.

In fig.~\ref{CNT_fig1} (body) we plot $V_{\mathrm{bg}}$-dependence of the linear response conductance of the left SWCNT QD. Within a wide range of $V_{\mathrm{bg}}$ we observe pronounced Coulomb blockade oscillations of irregular amplitude and spacing between neighboring peaks. At slightly positive $V_{\mathrm{bg}}$ in the range $1\,{\rm V}<V_{\mathrm{bg}}<3\,{\rm V}$ the oscillations are suppressed, indicating the energy gap in the spectrum characteristic for semiconducting SWCNTs. Depending on $V_{\mathrm{bg}}$ ranges, both oscillations with and without well-developed Coulomb blockade are seen. For some well-developed oscillations (ratio between conductance maxima and minima at least 10) we performed standard finite bias spectroscopy and found the addition energy, which varies between 1\,meV and 5\,meV. This scatter is not expected for an individual SWCNT QD, where the fluctuations of single-particle level spacing are typically much smaller than Coulomb energy, determined by the QD capacitance~\cite{Bockrath1997,Tans1997}. While it is difficult to draw a definitive conclusion, this might be another indication of a bundle of few SWCNTs connected in parallel. We also cannot exclude an impact of multi-subband transport at large negative $V_{\mathrm{bg}}$. In this case, transport irregularities can be related to a complex interplay of single-particle energy spectrum with intra-SWCNT and inter-SWCNT Coulomb interactions. Below we concentrate on current noise measurements within an arbitrarily chosen pair of adjacent Coulomb blockade resonances and correlate different transport regimes with the current noise measurements.

\begin{figure}
	\centering
	\includegraphics[width=\columnwidth]{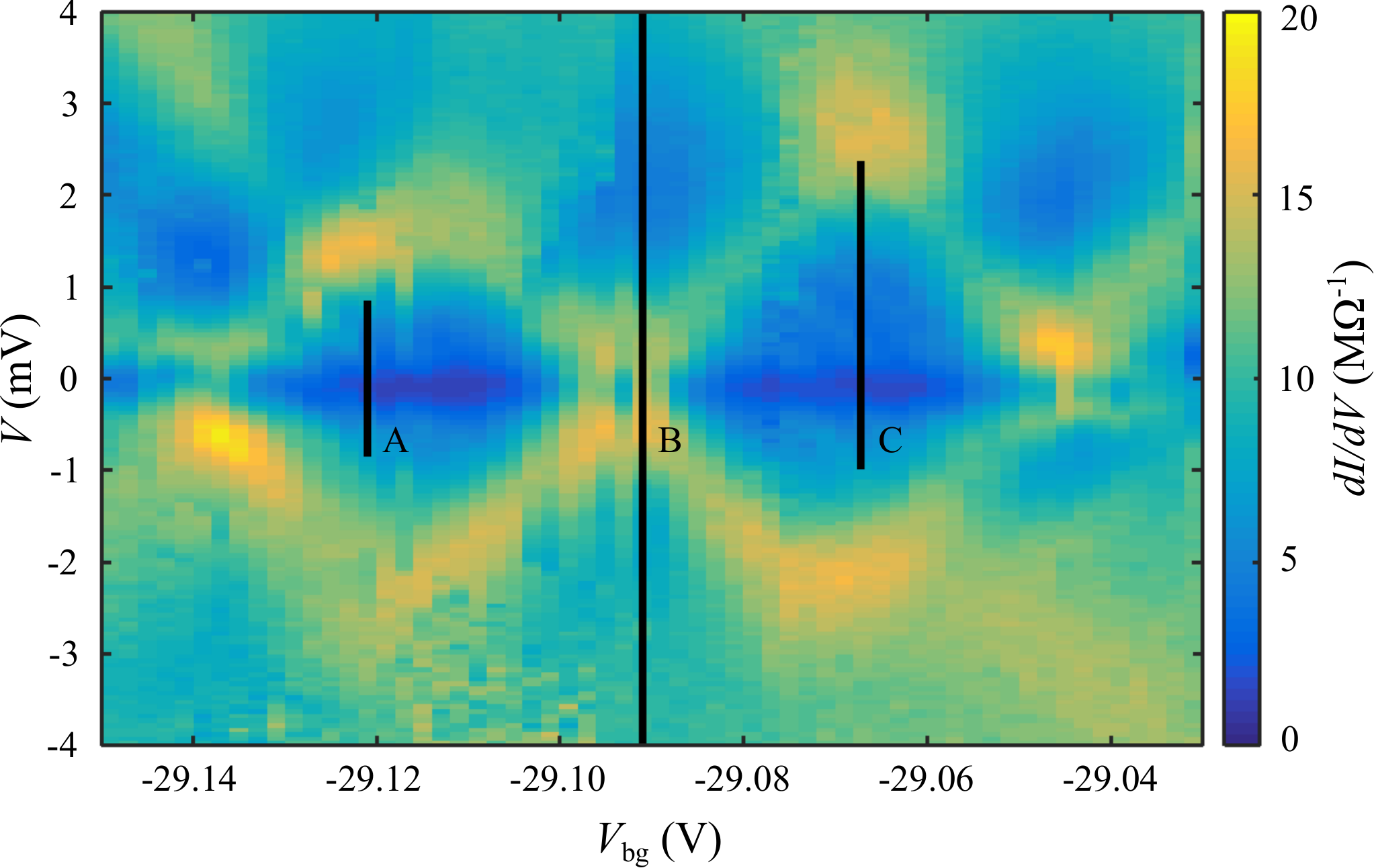}
		\caption{Differential conductance as a function of back-gate voltage $V_{\mathrm{bg}}$ and bias voltage $V$. For three $V_{\mathrm{bg}}$ values (marked by solid lines) conductance and noise are represented as a function of bias current $I$ and bias voltage $V$ in fig.~\ref{CNT_fig3} below. Here, the solid lines span only the intervals of $V$, corresponding to the Fano factor $F\approx1$ (A, C) and $F \approx 0.5$ (B).}\label{CNT_fig2}
\end{figure}

\begin{figure}
	\centering
	\includegraphics[width=1\columnwidth]{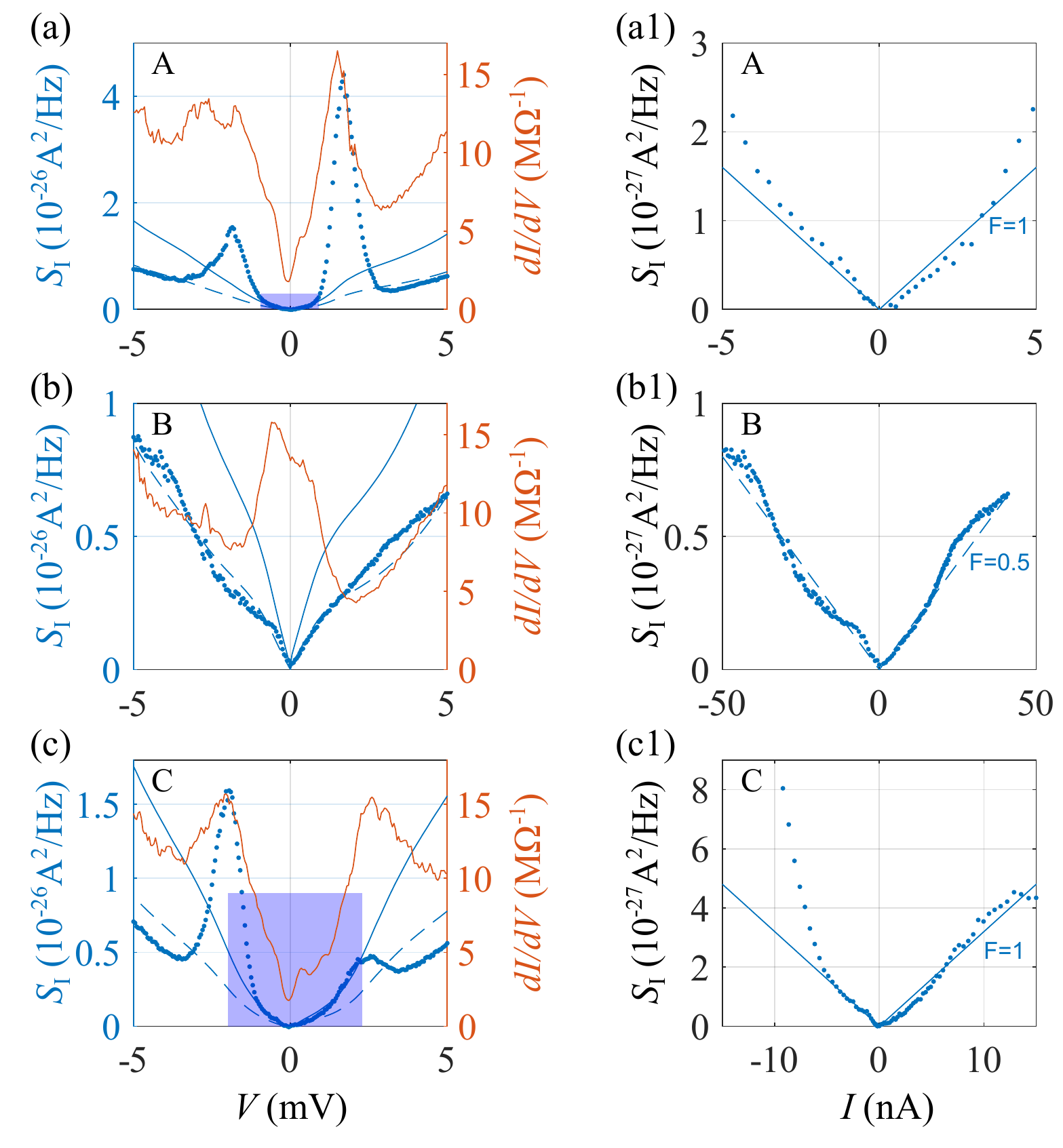}
		\caption{Noise and conductance along three representative cuts. (a, b, c) Current noise spectral density $S_I$ (blue dots, left axis) and differential conductance $dI/dV$  (orange curve, right axis) as a function of bias voltage $V$ for three values of $V_{\mathrm{bg}}$ corresponding, respectively, to the cuts A, B and C in fig.~\ref{CNT_fig2}. (a1, b1, c1) $S_I$ as a function of $I$ for the same data as in panels a, b and c. In panels a1 and b1 the data are plotted in the $I$ range, corresponding to shaded regions in panels a and b. Where applicable, blue guide lines indicate the current noise corresponding to the fixed Fano factors $F=1$ (solid lines) and $F=0.5$ (dashed lines).}\label{CNT_fig3}
\end{figure}

Fig.~\ref{CNT_fig2} shows a standard color-scale plot of differential conductance as a function of bias voltage $V$ and $V_{\mathrm{bg}}$. Note that owing to appreciable temporal drifts in our device this data cannot be directly compared to the data of fig.~\ref{CNT_fig1}, which were recorded more than a week before. The data of fig.~\ref{CNT_fig2} demonstrates suppressed conductance inside diamond-shaped Coulomb blockade regions. Here, sequential tunneling is forbidden and finite current through the QD occurs only via higher-order co-tunneling processes, which conserve the charge on the QD~\cite{Sukhorukov2001}. Figs.~\ref{CNT_fig3}a, \ref{CNT_fig3}b and \,\ref{CNT_fig3}c, axes on the rhs, show three traces of the differential conductance, $dI/dV$, as a function of $V$ obtained for fixed $V_{\mathrm{bg}}$ cuts of the Coulomb diamonds (marked by solid lines in fig.~\ref{CNT_fig2}). Traces A and C demonstrate wide conductance minima inside the Coulomb blockade region, followed by conductance maxima at bias voltages roughly corresponding to the boundaries of the Coulomb diamond structure. As usual, this behavior signals the transition from high order co-tunneling processes at low biases to sequential tunneling at high biases. By contrast, trace~B exhibits conductance maximum around $V=0$, indicating that charge transport near the Coulomb resonance occurs mainly owing to the sequential tunneling via a single quantum level of the SWCNT QD.  

The current noise brings more information about microscopic processes underlying SWCNT QD transport. Figs.~\ref{CNT_fig3}a, \ref{CNT_fig3}b and \ref{CNT_fig3}c, axes on the left, show the spectral density of current fluctuations measured for the same set of $V_{\mathrm{bg}}$. Depending on the gate voltage position, three distinct types of behavior are observed. First, at small currents and well inside the Coulomb diamonds, see the panels corresponding to traces A and C, the measured noise is close to Poissonian value $S_I\approx2eI$, as shown by solid guide lines. This behavior is expected for the shot noise of elastic co-tunneling process, during which the QD state remains in its ground state~\cite{Sukhorukov2001,Belzig2005,Basset2012,Harabula2018}. Qualitatively, in this case, the electron sees the QD as a single extremely opaque tunnel barrier, just like in a usual non-interacting case~\cite{Sukhorukov2001}. For convenience, we also replotted the low bias regions highlighted in figs.~\ref{CNT_fig3}a and\,\ref{CNT_fig3}c, in the form of $S_I$ vs $I$, respectively, in {\color{black} panels (a1) and (c1)
of fig.~\ref{CNT_fig3}}. In this representation the conventional linear shot noise behavior with $F\approx1$ is evident.  

Second, when the QD chemical potential is tuned to the position of Coulomb resonance, see the {\color{black} panels (b) and (b1)
of fig.~\ref{CNT_fig3}} corresponding to the trace B, the noise is reduced compared to the Poissonian value. The reduction is roughly by a factor of 2 and slightly depends on current, see the dashed guide line with $F=0.5$. The observation of $0.5\leq F\leq1$ is a signature of sequential tunneling in which transport through the QD occurs via two independent tunneling processes across the QD-lead barriers~\cite{Sukhorukov2001,Belzig2005,Onac2006,Wu2007,Basset2012,Harabula2018}. Similar to the non-interacting double-barrier case, this effect is a consequence of current conservation~\cite{Blanter2000,Sukhorukov2001} and $F\approx0.5$ indicates that QD-lead barriers are nearly identical. Close to linear $S_I(I)$ dependence in this case is evident from {\color{black} panel (b1)}. Passing, we note that at some $V_{\mathrm{bg}}$ at small currents we have obtained $0.3<F<0.5$, which is difficult to understand in a conventional framework of shot noise in the regime of sequential tunneling. In these situations, however, the {\color{black} SWCNT} resistance was typically below 100 k$\rm\Omega$ and clear Coulomb diamond structure was not observable. 

Third, and perhaps the most striking, is the observation of super-Poissonian noise peaks, sometimes with $F\sim8$, around certain values of $V$ and $V_{\mathrm{bg}}$, see the data for traces A and C near $V\sim\pm2\,$mV. Although much more pronounced, this effect resembles super-Poissonian noise maxima observed in random hopping transport via interacting localized states in the insulating phase of n-GaAs transistor~\cite{Safonov2003}. In {\color{black} carbon-nanotube} QDs the shot noise several times the Poissonian value was measured in the regime of inelastic co-tunneling~\cite{Onac2006,Harabula2018}. In general, such a behavior is explained by modulation of the QD current, which occurs owing to the very fast random switching between different quantum states in the regime of Coulomb blockade~\cite{Sukhorukov2001,Belzig2005,Kaasbjerg2015}. Giant noise then is a pure interaction effect not observed, e.g., in SWCNTs in a Fabry-Perot transport regime~\cite{Wu2007}. In Markovian approximation~\cite{kogan,Sukhorukov2001} the modulation noise acquires Lorentzian spectrum and covers the frequency range up to the inverse correlation time of the switching process. Measurements in SWCNTs in the gigahertz regime indicate that the correlation time falls in the range of a few 10\,ps~\cite{Onac2006,Basset2012}, which effectively gives rise to super-Poissonian white noise at sub-10\,GHz frequencies~\cite{Harabula2018}, including those used in the present experiment. 

As seen from figs.~\ref{CNT_fig3}a and~\ref{CNT_fig3}c, super-Poissonian noise maxima correlate with the maxima of differential conductance, which is a clear signature of the modulation noise. Apart from that, however, the low quality of the Coulomb diamonds  in our experiment, see fig.~\ref{CNT_fig2}, precludes us from unequivocal identification of the microscopic SWCNT QD states participating in the switching process.

%\begin{figure}
	%\centering
	%\includegraphics[width=\columnwidth]{map2_w}
	%\caption{Top panel: conductance as a function of backgate voltage $V_{backgate}$ and bias voltage $V_{SD}$. Bottom panel: current noise spectral density divided by the Schottky noise $|2eI|$. For several $V_{backgate}$ values (red dashed lines) conductance and noise are represented as a function of current on the next figure.}
%\end{figure}
%
%\begin{figure}
	%\centering
	%\includegraphics[width=\columnwidth]{graphs2_w}
	%\caption{Current noise spectral density $S_I$ (blue dots) and conductance (orange curve) $G$ as a function of current for 4 different values of $V_{backgate}$. A and D: high noise ($S_I>2eI$) and negative conductance; B: quantum dot is in the "open" state; C: "closed" state with high noise ($S_I>2eI$). The blue lines represent shot noise for a different Fano factor $F$.}
%\end{figure}
%

\section{V. Noise and fluctuations of temperature at a resistive transition}

Spontaneous fluctuations of current in a conductor at the onset of the transition to the superconducting state (called below the resistive transition) are believed to provide microscopic insight into superconducting correlations. More than 50 years ago, giant low-frequency noise in type-II superconducting films in the magnetic field was associated with the correlated motion of bunches of magnetic vortices~\cite{VanOoijen1965}. Much smaller noise in disordered type-I films in zero field  was interpreted in terms of the motion of independent vortices, emerging from unbinding of vortex-antivortex pairs in a 2D superconductor~\cite{Voss1980}. Giant noise can also be observed in superconducting weak links~\cite{Hoss2000}, originating from fluctuations of the order-parameter~\cite{Bagrets2014}, multiple~\cite{Zonda2015} or individual quantum~\cite{Semenov2016} phase slips. In the context of transition edge thermometry~\cite{Hoevers2000}, it was argued that giant noise at the resistive transition can also originate from spontaneous fluctuations of the electronic temperature. Here we present the ab-initio estimate of such noise, {\color{black} which} is insensitive to the microscopic features of the resistive transition and can dominate the noise spectrum in a wide frequency range. 

In thermodynamic equilibrium mean squared fluctuation of the electronic temperature, $T_e$, is expressed as~\cite{landauV}:
\begin{equation}
	\left\langle \Delta T_e^2\right\rangle = \frac{k_BT_e^2}{C},
	\label{dT_eq1}
\end{equation}
where $C$ is the (electronic) heat capacity and $k_B$ is the Boltzmann constant. We are interested in the spectral density of fluctuations as a function of frequency, $S_T(f)$, which obeys the standard identity $\left\langle \delta T_e^2\right\rangle=\int S_T(f)df$. Note, that $S_T(f)$ is twice the Fourier transform of the correlation function~\cite{kogan} and can be obtained, e.g., in ac-heating experiments~\cite{Kogan1985}. Typically, such experiments identify Lorentzian-shaped frequency response~\cite{Kardakova2016}, with characteristic correlation time of a temperature fluctuation, $\tau$, given by the ratio of $C$ and the heat conduction rate to the external bath, $G$, (both per unit volume), $\tau=C/G$. Below we assume that heat conduction is determined by the electron-phonon (e-ph) cooling rate, $G=G_{\mathrm{e-ph}}$, and derive from eq.~(\ref{dT_eq1}) $S_T(0)=k_BT^2/G_{\mathrm{e-ph}}$, omitting {\color{black} an insignificant for our purposes} numerical factor of order unity.

Fluctuations of $T_e$ give rise to fluctuations of the resistance, $R$, and hence to voltage fluctuations in the current biased device. These fluctuations are strongest at the resistive transition, where $R(T_e)$ has the strongest temperature dependence. 
The spectral density of such voltage fluctuations is given by~\cite{Kogan1985} $S_V=I^2S_R$, where $I$ is the average current and $S_R$ is the spectral density of the resistance fluctuations. Since $S_R=\left(\frac{dR}{dT}\right)^2S_T$, we obtain:
 
\begin{equation}
	S_V(0)=G_{\mathrm{e-ph}}^{-1}I^2\left(\frac{dR}{dT}\right)^2k_BT^2, \label{eq3}
\end{equation}
where, $G_{\mathrm{e-ph}}$ is the total e-ph heat conduction of the sample. 

In order to estimate the magnitude of voltage fluctuations, we assume that the current biased superconducting film in the vicinity of the sharp resistive transition remains in local thermal equilibrium. In other words, we assume that while Joule heating slightly raises the electronic temperature above the bath, $\delta T_e\equiv T_e-T\ll T$, all the electronic degrees of freedom are mutually equilibrated and the $I-V$ characteristics of the device simply follows the temperature dependence of the resistance $V/I=R(T_e)$. {\color{black} This assumption is reasonable at least in some cases, see Ref.~\cite{Postolova2015}}. Since $\delta T_e$ is related to the e-ph cooling rate $\delta T_e=(IU)G_{e-ph}^{-1}$, we obtain:

\begin{equation}
	S_V(0) = k_B\delta T_e\frac{T^2}{R(T_e)}\left(\frac{dR(T_e)}{dT_e}\right)^2 \label{dT_eq4}
\end{equation}
 
Under our assumptions, the spectral density of voltage fluctuations is simply determined by the temperature dependence of the resistive transition $R(T_e)$ with $T_e$ as a parameter. Obviously, noise predicted by eq.~(\ref{dT_eq4}) can very well exceed the equilibrium Johnson-Nyquist value $S_V/(4k_BTR)\sim T/4\Delta T\gg 1$, provided the width of the resistive transition, $\Delta T$, is small enough. It's also straightforward to see that for high-quality films with $R$ in the range of a few $\rm k\Omega$, our estimate of $S_V$ is much higher than the noises caused by quantum phase slips~\cite{Semenov2016} or by fluctuations of the order-parameter~\cite{Bagrets2014}. Unlike these mechanisms, however, the noise owing to thermal fluctuations is insensitive to the microscopic nature of the resistive transition and its spectral density is cutoff at a frequency on the order of the inverse e-ph relaxation time~\cite{Kardakova2016}.

\section{Summary}

In summary, we discussed several examples how non-equilibrium noise measurements shed light on microscopic aspects of mesoscopic electron transport. Such experiments directly probe electronic correlations, elastic scattering, and energy relaxation in various transport regimes, from normal to superconducting and from ballistic to localized. Hopefully, in this short review, we demonstrated that measuring noise is not only a powerful but also a beautiful approach in experimental condensed matter physics.

%\begin{figure}[t]
%\begin{center}
%%\vspace{5mm}
%\hspace*{-5mm}
  %\includegraphics[width=1\linewidth]{?.png}
%\end{center}
  %\caption{} 
	%\label{}
%\end{figure}

We are grateful to Z.D.~Kvon, J.~Becker and D.~Ruhstorfer for help and acknowledge discussions with K.E.~Nagaev, T.M.~Klapwijk, A.V.~Semenov, A.G.~Semenov, M.A.~Skvotsov, G.E.~Fedorov and A.A.~Zhukov. The experimental work was partly supported by the RSF Project No. 16-42-01050 and theoretical estimation in section V was supported by the RSF Project No. 17-72-30036. G.K. acknowledges support from DFG-Project No. KO-4005/5-1.

\bibliography{summary_bib}

\end{document}